\documentclass[11pt]{article}
\usepackage{latexsym}

\usepackage{amssymb}
\usepackage{amsmath}

 \usepackage{epsfig}
 \usepackage{graphicx}

\newcommand{\be}{\begin{equation}}
\newcommand{\ee}{\end{equation}}
\newcommand{\ba}{\begin{eqnarray}}
\newcommand{\ea}{\end{eqnarray}}
\newcommand{\baa}{\begin{array}}
\newcommand{\eaa}{\end{array}}
\newcommand{\bi}{\begin{itemize}}
\newcommand{\ei}{\end{itemize}}
\newcommand{\edoc}{\end{document}}

\newcommand{\la}[1]{\label{#1}}

\def\gsim{\raise0.3ex\hbox{$>$\kern-0.75em\raise-1.1ex\hbox{$\sim$}}}
\def\lsim{\raise0.3ex\hbox{$<$\kern-0.75em\raise-1.1ex\hbox{$\sim$}}}

\begin{document}
\begin{center}
\Large
Quark Matter '84, Helsinki, 17-21 June 1984

\large
K.  Kajantie

Helsinki Institute of Physics, P.O.Box 64, FI-00014 University of Helsinki, Finland
\end{center}

\normalsize

\section{Background of QM84}

To dig deep into the origins of Quark Matter 1984 in Helsinki one must start from early seventies
when Helmut Satz visited us 
in November 1970 - August 1971. In May 1971 we (the organising committee was Byckling, Kajantie,
Satz and Tuominiemi) organised in Helsinki a "Colloquium on Multiparticle Dynamics" \cite{ismd2}, which was
declared to be the 2nd meeting in a new series to be\footnote{These meetings and their proceedings
can simplest be found by going to cds.cern.ch and searching under Proceedings for, say, ``multiparticle"}, the 1st one was at Ecole Polytechnique, Paris,  
in May 1970 \cite{ismd1}. And a series it has become: the 45th Symposium
on Multiparticle Dynamics ISMD 2015 was organised this year. The highlight was Alan Krisch of Argonne rushing 
from CERN to Helsinki to show the first results on inclusive pion distributions from the Intersecting
Storage Ring. These extended up to then impressive $p_T$ value of almost 1 GeV. One also
learnt that pions are dominantly produced at 90 degrees in the center of mass system.
No quarks or gluons entered the discussion: QCD was just in the process of being invented.
Nevertheless, this was a precursor of the type of physics at QM meetings. 

Even after QCD was invented around 1972 statistical models were a source of inspiration for
understanding the dynamics of hadronic collisions. Again Helmut was active and organised a
Bielefeld Summer Institute on Many Degrees of Freedom in Particle Theory in
August-September 1976  \cite{satz1978}. There were lots of seminars on pre-QCD theory,
solitons, Reggeons, dual models, etc, but also quark matter related ones like neutron
stars and lattice QCD. Perhaps characteristic of the times is that there also was a
general discussion, moderated by A. Krzywicki, on "Is there a connection between Field
Theory and Phenomenology". One was slowly absorbing the idea that QCD really would
describe hadronic experiments.  My talk on dilepton production via the Drell-Yan mechanism could also
be counted among those related to quark matter. In fact, next summer 1977 I was the plenary
speaker in the European High Energy meeting in Budapest on Drell-Yan production of dileptons,
which was hailed as evidence of QCD degrees of freedom in hadronic processes. I had a long
argument with Helmut about this interpretation: he still preferred a statistical model for production of
dilepton pairs.

At the end of seventies QCD and
asymptotic freedom were known, the idea of deconfined quark-gluon matter was in the air and lots of
papers had already been written. Helmut Satz went through the literature and invited everybody even
remotely connected with the topic to a meeting Statistical Mechanics of Quarks and Hadrons  in 1980  \cite{satz1981}. 
This Bielefeld meeting was a real milestone and 
should in my opinion be counted as the 1st meeting in the QM series. It was genuinely a starting point --
theoretical of course since there were no ultrarelativistic experiments, only ideas for them. 
One of the participants was Larry
McLerran with ideas about perturbative QCD at  $T=0$ and about  lattice Monte Carlo at finite $T$.
He then visited Helsinki from April to July 1982.

To organise a meeting one needs an ecosystem of researchers. This developed rapidly 
in early eighties also in Helsinki, 
the field was wide open for new ideas. On the basis of what I learnt in  \cite{satz1981} I,
with Montonen and Pietarinen, extended the SU(2)
thermal lattice Monte Carlo work by McLerran and Svetitsky and others to SU(3) in March 1981.
This is actually the first time SU(3) was simulated.
My first work on dilepton production in nuclear collisions with Hannu I. Miettinen was submitted also in March 1981. At the
end of 1981 I worked at CERN with Joe Kapusta on thermal perturbation theory. Catalysed
by joint hydro work with McLerran in April-July 1982 I submitted with Raitio what I believe are the first 
numerical  computations of ultrarelativistic hydrodynamical flow in November 1982.  This
hydrodynamics project was joined by Vesa Ruuskanen in 1983. The work on deflagrations and
detonations with Gyulassy, Kurki-Suonio and McLerran, submitted in July 1983 had been very
important if nature had been kind enough to provide us with a first order transition. 
Altogether, we had in Helsinki a group willing and able to undertake organising
a conference. 

Satz organised a further meeting on QCD matter formation and heavy ion collisions 
in Bielefeld on 10-14 May 1982 \cite{satz1982i}. 
The title indicates that thinking of experimental observation of quark matter had started. This is nowadays
counted as the 2nd meeting in the Quark Matter series.  
With the rather extensive participation of Finnish physicists it was not unexpected that
one planned to have the following meeting in the series  in Helsinki.
However, the enthusiasm for the physics was such, especially in the US, that 
Brookhaven Nationa Laboratory (BNL)
wanted to organise the next one. In the US the closure of the p+p collider
project Isabelle  was imminent (finally decided in July 1983) and it was obvious that the infrastructure
already built for it could be utilised by building an A+A collider. To prepare the ground for it  BNL
organised Quark Matter number 3,  in 1983. It was an enormous success, 250 participants
was more than one could expect. The final outcome was the approval of the 
Relativistic Heavy Ion Collider project in 1991.

\section{QM84 in Helsinki}
The first meeting of the organising committee took place on 12 October 1982. Those present
were Kajantie (chair), Miettinen, Montonen, Ruuskanen and Toimela (secretary), all authors of
known papers in the physics of QCD matter. Toimela left soon to prepare his thesis and from
October 1983 Jukka Maalampi, now professor in Jyv\"askyl\"a,  became the secretary. Also
Risto Raitio joined as a committee member.

The title of the meeting evolved in the course of organisation: first it was 4th International Symposium on the 
Physics of Quark Gluon Plasmas but the final form was 4th International conference on
ultra-relativistic heavy ion collisions. This reflects the increasing importance of experiments. 
The compact practical name Quark Matter '84 was introduced following BNL. 
One can observe that the organisation started in good time, the dates of the meeting were fixed to be
Sunday 17 - Thursday 21 June 1984.

The organisation progressed smoothly. The venue was chosen to be the Swedish-Finnish cultural
center on a small island next to Helsinki (Oct 82), money was gathered from Finland and
Nordita, Copenhagen, advisory committee was
appointed (Oct 83), Springer was chosen as proceedings publisher \cite{kajantie}, program was outlined
and invitations were sent (Nov 83),
1st bulletin was mailed in Feb 84 and the 2nd one in Apr 84. Communication was carried out
by using letters, telexes and telephone - no email yet. But \TeX\, had started its spread for
the contributions to the proceedings.

The venue was a success - but for the weather. One major mishap though: Gordon Baym
arrived in the middle of the night with other Americans and found that there was no room for
him at the hotel. For sure a reservation had been made. Fortunately a room was found at a distant hotel
for this seasoned traveler.

The number of participants was 100, roughly 40 experimentalists and 60 theorists. This 
was a sizable reduction from the 250 at BNL the year before. One might say this had to do with
the distance but not: out of the participants 43 came from the US. This simply reflects
the enormous enthusiasm generated by building a heavy ion collider at BNL. What is also 
interesting is that out of the 100 in Helsinki the following 13 also participated in QM2015  31 years later: 
Baym, Busza, Friman, Gavai, Gyulassy, Kapusta, McLerran, B. M\"uller, Nagamiya, Otterlund, Sumbera, 
Tannenbaum, Zajc. The field of research was new and expanding, attracted young people who were
occupied by its problems for decades to come.

The 100 participants gave 21 plenary (included in the proceedings \cite{kajantie}) and 44
parallel talks (only listed in the proceedings). What is striking is that the general outline of what one would see in and
accomplish with ultra-relativistic heavy ion collisions was known already then: the intervening
three decades have only brought an enormous amount of beautiful experimental and
theoretical detail to this general framework. Is there at present anything similarly novel and transformative
for the future in our branch of physics?

\begin{figure}[!t]
\begin{center}


\includegraphics[width=0.7\textwidth]{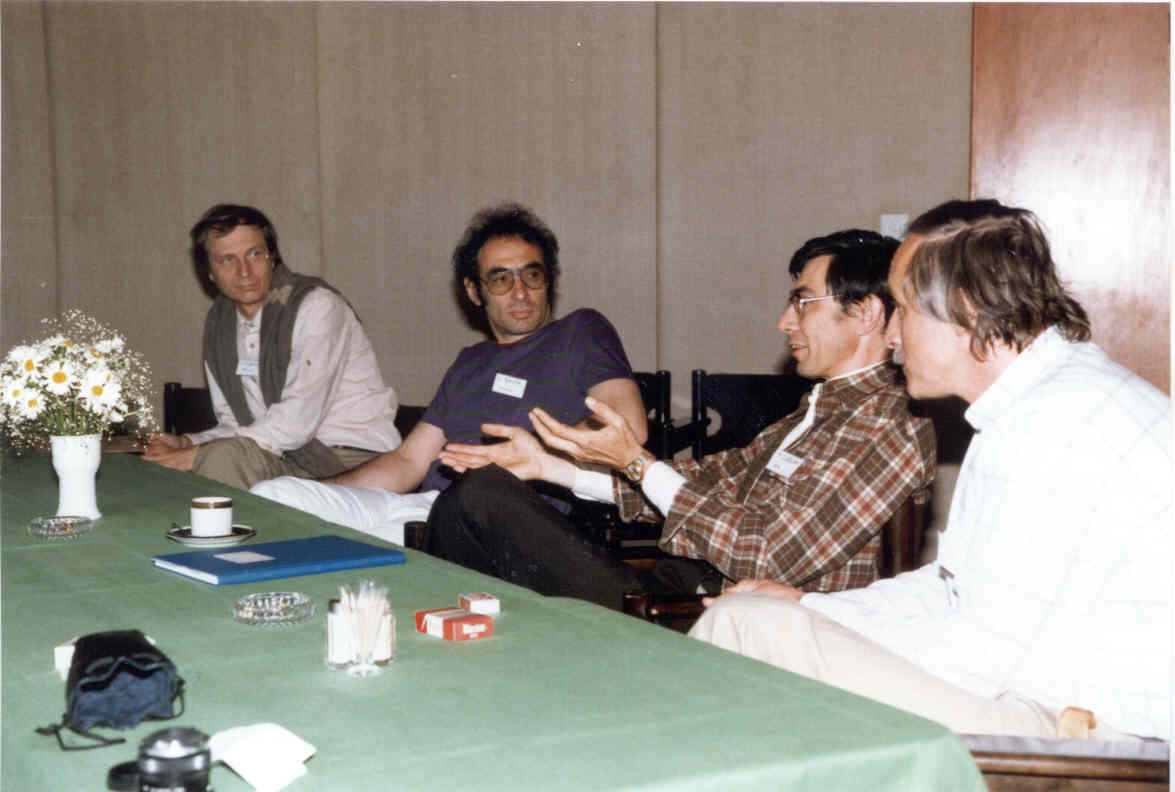}
\end{center}

\caption{\small K.Kajantie, G.Baym, T.Ludlam and R.Stock at the press conference  }
\la{Xh}
\end{figure}

\begin{figure}[!t]
\begin{center}


\includegraphics[width=0.32\textwidth]{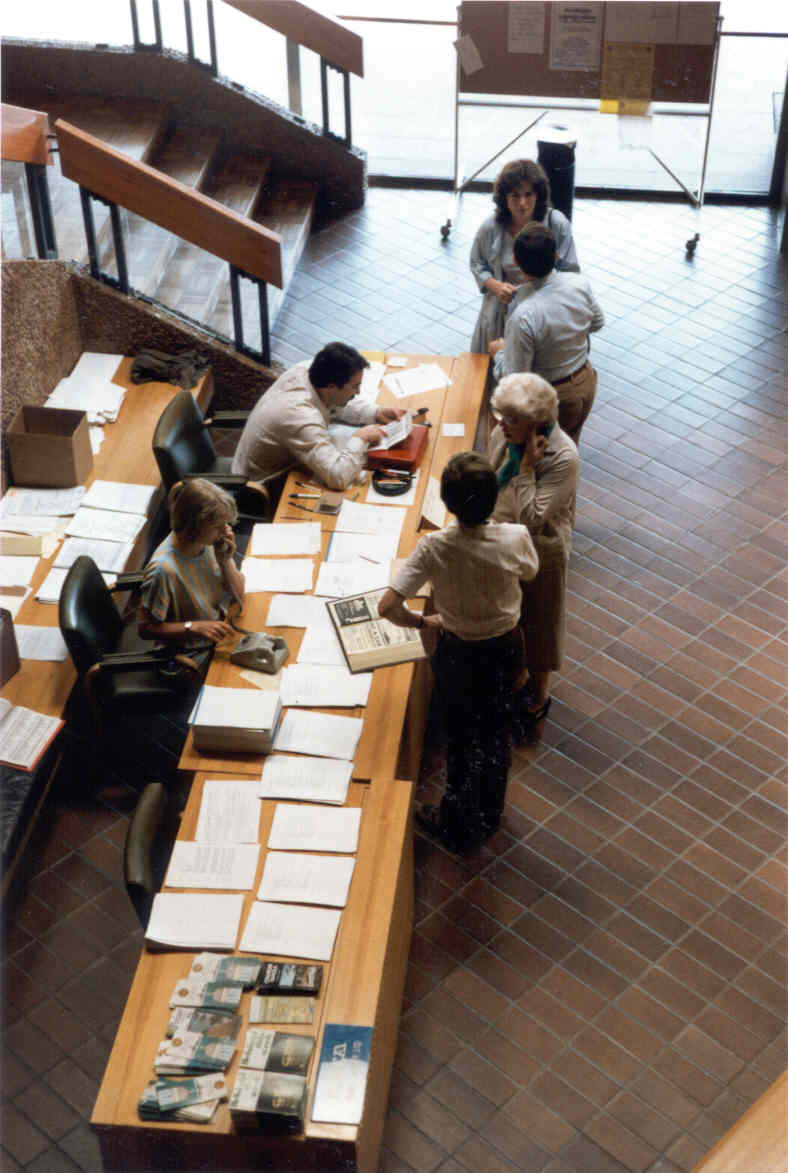}\hfill
\includegraphics[width=0.32\textwidth]{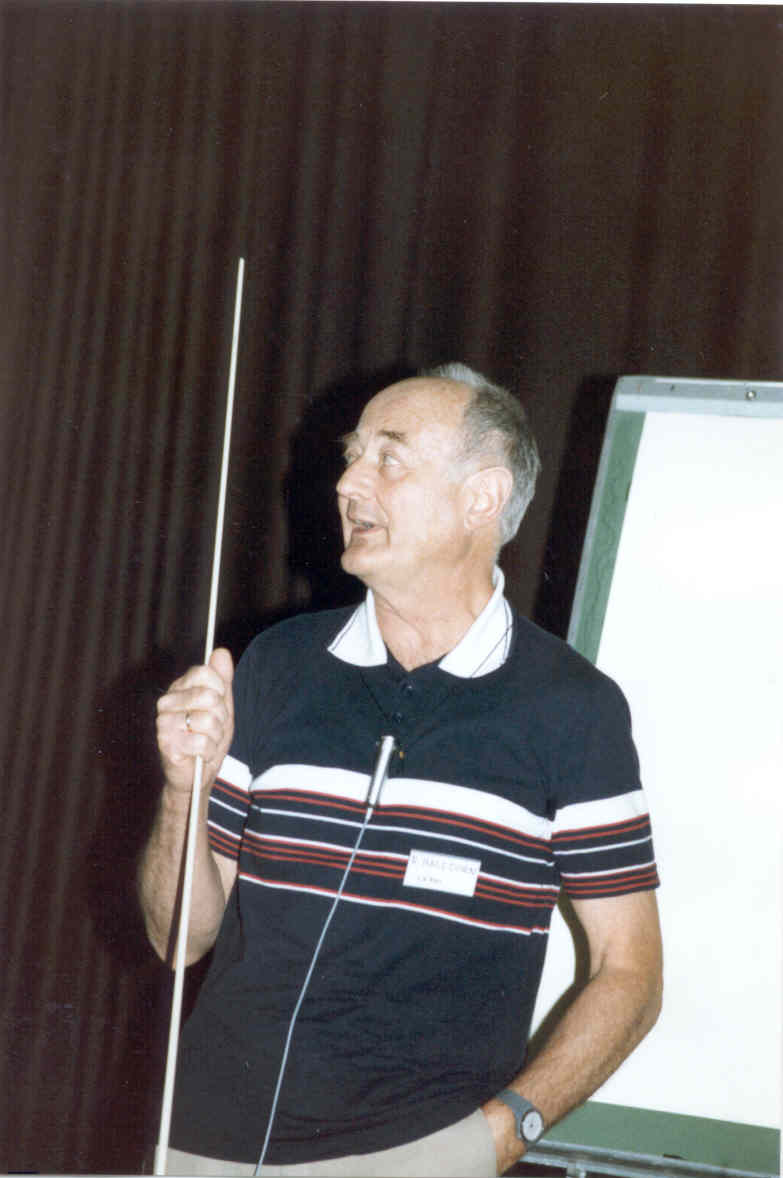}\hfill
\includegraphics[width=0.32\textwidth]{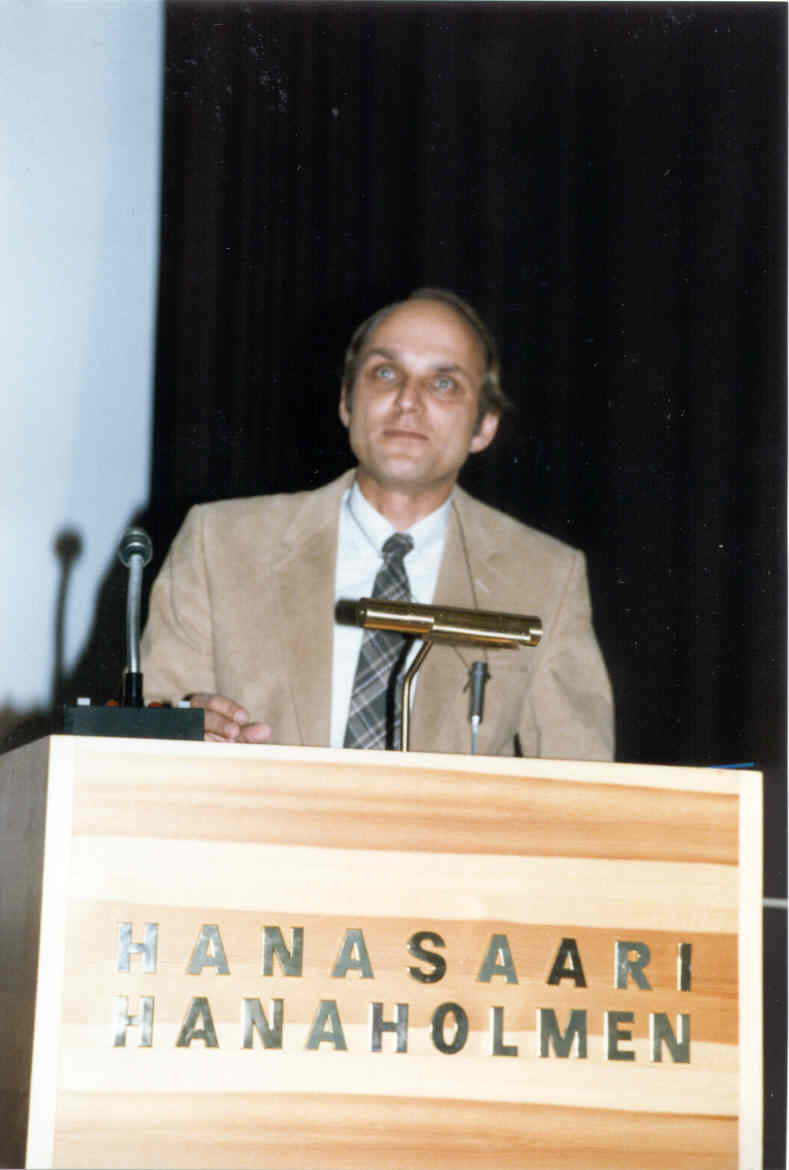}
\end{center}

\caption{\small Registration, Rolf Hagedorn speaking and Helmut Satz giving the summary talk.  }
\la{Xh1}
\end{figure}

\end{document}